\documentclass[twocolumn,showpacs,prb,floatfix,amssymb,aps]{revtex4}
\usepackage{graphicx}
\usepackage{color}

\begin{document}

\hsize\textwidth\columnwidth\hsize\csname@twocolumnfalse\endcsname

\title{Ferromagnetism in a two-component Bose-Hubbard model with a synthetic spin-orbit coupling}
\author{Jize Zhao$^{1,2}$}\email{jizezhao@gmail.com}
\author{Shijie Hu$^3$, Jun Chang$^4$, Ping Zhang$^{1,2}$}
\author{Xiaoqun Wang$^{5,6}$}\email{xiaoqunwang@ruc.edu.cn}
\affiliation{$^1$LCP, Institute of Applied Physics and Computational Mathematics, Beijing 100088, China}
\affiliation{$^2$Beijing Computational Science Research Center, Beijing 100084, China}
\affiliation{$^3$Max-Plank-Institut f\"ur Physik komplexer Systeme, Dresden 01187,  Germany}
\affiliation{$^4$Institute of Theoretical Physics and Kavli Institute for Theoretical Physics, CAS, Beijing 100190, China}
\affiliation{$^5$Department of Physics and Beijing Laboratory of Opto-electronic Functional Materials $\&$ Micro-nano Devices, Renmin University of China, Beijing 100872, China}
\affiliation{$^6$Department of Physics and Astronomy, Shanghai Jiao Tong University, Shanghai 200240, China}
\begin{abstract}
  We study the effect of the synthetic spin-orbit coupling in a two-component Bose-Hubbard model in one dimension by employing the density-matrix renormalization group method. A ferromagnetic long-range order emerges in both Mott insulator and superfluid phases resulting from the spontaneous breaking of the $Z_2$ symmetry, when the spin-orbit coupling term becomes comparable to the hopping kinetic energy and the intercomponent interaction is smaller than the intracomponent as well. This effect is expected to be detectable with the present realization of the synthetic spin-orbit coupling in experiments.
\end{abstract}

\maketitle

\section{Introduction}

Since the first experimental realization of Bose-Einstein condensation (BEC) in dilute atomic gas in 1995 \cite{ANDERSON1, DAVIS1},
a research area of ultracold atomic systems has opened and rapidly expanded to interplay with other physics.
In particular, optical-lattice systems of ultracold atoms has become a very powerful platform for systematical investigations
on the nature of novel ground states and quantum phase transitions for various strongly correlated bosonic and fermionic
systems. The advantages over solids such as controllability and purity
allow us to explore various fundamental physics of strongly interacting tight-binding models associated with the long-standing issues of condensed matter physics. Appealing examples are Bose-Hubbard model \cite{FISHER1} and its generalizations \cite{DUAN1, FUKUHARA1}.
The Mott-superfluid transition predicted theoretically  has been experimentally observed \cite{GREINER1}. More recently,
a synthetic spin-orbit (SO) coupling has been successfully engineered in BEC of an ultracold atomic system \cite{LIN1}.
The capability of the platform is thus greatly enhanced to simulate more complicated correlated systems with SO couplings.

Relativistic SO couplings can be crucial for the presence of unusual magnetic properties in solids, where the
well-known Dzyaloshinsky-Moriya (DM) interaction \cite{DZ,MORIYA} can be derived to explain different
magnetic ground states and excitations \cite{Broholm,Oshikawa,Zhao}. In recently discovered topological
insulator \cite{KANE1, QI1}, a SO coupling is substantial for understanding gapless topological edge states,
similar to the edge states in quantum Hall effects, but without breaking time-reversal symmetry. Although
a increasing number of new phenomena are seemingly related to SO couplings, full investigations on SO coupling
effects are hindered by the fact that SO couplings can seldom be manipulated artificially
in solids. In ultracold atom systems \cite{LIN1, WANG1, CHEUK1}, however, the SO coupling becomes adjustable
since it is engineered with a pair of tunable lasers via creating momentum-dependent interaction between
two internal atomic states. This vigorously activates intensive investigations on intriguing SO coupling
effects for various correlated bosonic and fermionic systems \cite{HU4, WU1, ZHOU1, DENG1, ZHANG1, QU1, ZHAI1}.

In the presence of optical lattices, several theoretical groups have studied phase diagrams of
the two-component bosonic Hubbard model mainly with Rashba SO couplings in two dimensions and with various
approximations\cite{MANDAL1,CAI1,RADIC1,GONG1,COLE}. However, at present, the realizable SO coupling in
experiments is along one direction. It seems that little attention has been paid to the experimentally realizable one because
it can be eliminated by a gauge transformation.
However, in the presence of spin-dependent interaction, such a SO coupling can not be simply gauged away and
we found that a ferromagnetic phase can be induced in both Mott insulator (MI) and superfluid (SF) phases
when the intercomponent interaction is smaller than the intracomponent one.

In Sec. II, we introduce the Hamiltonian that we study; in Sec. III, we present the phase diagrams.
Excitation gaps, correlation functions and entanglement entropy are discussed; in Sec. IV, we give
our conclusions.

\section{Hamiltonian}

Since the synthetic SO coupling currently realized in experiments is of the one-dimensional nature, we focus on a one-dimensional two-component Bose-Hubbard model with such a synthetic SO coupling term. Each component represents one hyperfine state of ultracold atoms. The Hamiltonian reads
\begin{eqnarray}
\hat{H} & = & -t\sum_{i\tau} \left(\hat c_{i\tau}^\dagger \hat c_{i+1\tau}+h.c.\right)
              +\hat{\mathcal{T}}_{so}
              +\frac{U}{2}\sum_{i\tau} \hat n_{i\tau}( \hat n_{i\tau}-1) \nonumber \\
        &   & + U^{\prime} \sum_i \hat n_{i\uparrow} \hat n_{i\downarrow}
              -\mu\sum_{i}(\hat n_{i\uparrow}+\hat n_{i\downarrow}).
 \label{HSO}
\end{eqnarray}
where $i$ indicates $i$-th lattice site and runs from $1$ to $L$, and a spin index $\tau$ is either $\uparrow$ or $\downarrow$ to denote two components of bosonic atoms. $\hat c_{i\tau}(\hat c^{\dagger}_{i\tau})$ is the annihilation (creation) operator at the site $i$ with a spin $\tau$, $t$ the hopping integral between the nearest neighbour sites and $\hat n_{i\tau} = \hat c^{\dagger}_{i\tau} \hat c_{i\tau}$ the boson number operator for site-$i$ with spin $\tau$. $U$ is for an on-site intracomponent interaction, while $U^{\prime}$ is the one for opposite spins, i.e. an inter-component interaction. $\mu$ is the chemical potential to control the filling factor of bosons. $\hat{\mathcal{T}}_{SO}$ represents the SO coupling term and takes a form of $\hat p_x\sigma^y$ which corresponds to the current experimental realization \cite{LIN1}, and $\sigma^y$ is the $y$-component of Pauli matrix. In a tight-binding form, we have
$\hat{\mathcal{T}}_{SO} = -\lambda\sum_{i}(\hat c^{\dagger}_{i\uparrow} \hat c_{i+1\downarrow} - \hat c^{\dagger}_{i\downarrow} \hat c_{i+1\uparrow})+h.c.$,
where $\lambda$ is the SO coupling strength.

In the isotropic interacting case, i.e. $U^\prime=U$, it has been shown that $\hat{\mathcal{T}}_{SO}$ can be eliminated by a site-dependent rotation in the internal space \cite{CAI1}. The Hamiltonian (\ref{HSO}) reduces to the standard two-component Bose-Hubbard model(TBHM), whose properties have been studied extensively in literature \cite{ALTMAN1, MISHRA1, HU3, ISACSSON1}. In this sense, the SO coupling is trivial in the isotropic interacting case. Thus, only in combination with a Zeeman term the SO coupling may lead to interesting physics. However, when $U^{\prime} \ne U$, $\hat \mathcal{T}_{SO}$ cannot be gauged away. Even in the absence of a Zeeman term, we found that new magnetic phases emerge from both MI and SF phases with  $U^{\prime} < U$. While the spiral magnetic phase \cite{CAI1} for $U^{\prime} = U$ is seemingly inherited for $U^{\prime} > U$, below we focus on the most unexpected case of $U^{\prime} <U$ and leave those discussions with $U^{\prime} > U$ elsewhere. For the convenience, we take $U^{\prime}=0.2U$ in the following discussions.

\section{Numerical results and discussions}

\subsection{Symmetry and phase diagrams}

To establish the phase diagram of the Hamiltonian (\ref{HSO}), here we employ density-matrix renormalization group (DMRG) \cite{WHITE1}. This is a quasi-exact numerical method which has been very successfully used to investigate various properties of quasi-one dimensional correlated systems \cite{PESCHEL, SCHOLLWOCK1}.
In our calculations, we impose an open boundary condition and restrict the filling factor to $n=1$. In the application to bosonic systems, it is necessary to truncate the Hilbert space for each site. For the present computation, the dimension of each component at each site is truncated to 4 so that the degree of freedom is 16 per site \cite{HU1}. 500 $\sim$ 1200 states are kept to ensure truncation errors are no larger than  $10^{-7}$. Moreover, we utilize an accurate finite-size algorithm with 2-6 sweeps at the chain length of $L=128$, to reach the convergence of the seventh digit for the ground state energy per site \cite{HU2}.

\begin{figure}[h]
\includegraphics[width=9cm, clip]{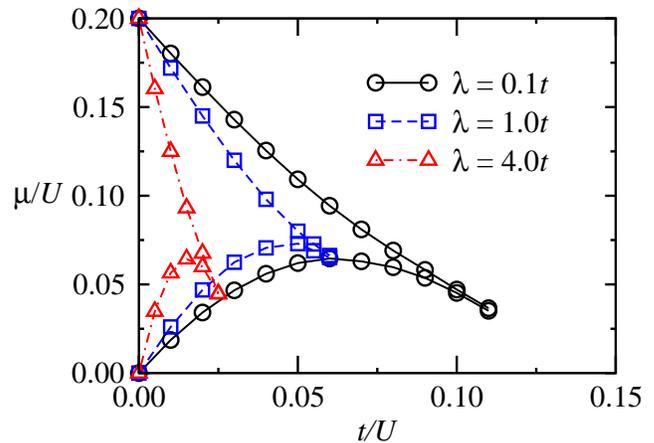}
\caption{(Color online) Phase diagrams with the first Mott lobe for $\lambda=0.1t, 1.0t, 4.0t$. For each $\lambda$, the region of the Mott-insulator phase is surrounded by the $\mu$-axis and a colorful curve connecting symbols, and the rest in the $t-\mu$ plane for the superfluid phase.}
\label{fig1}
\end{figure}

We first examine how the SO coupling affects the phase diagram for the MI-SF transition. Fig. \ref{fig1} shows a $t-\mu$ phase diagram with $\lambda=0.1t, 1.0t, 4.0t$. The upper and lower boundaries of MI phase are determined by chemical potentials \cite{KUHNER1} $\mu^+ = E_{0}(N+1,L)-E_{0}(N,L)$ and $\mu^-=E_{0}(N,L)-E_{0}(N-1,L)$, respectively, where $E_0(N, L)$ is the ground state energy with $N$ bosons for given $L$. MI is characterized by a finite single-particle excitation gap $\Delta_c=\mu^+ - \mu^-$.
We note that at $t=0, \lambda=0$, Hamiltonian (\ref{HSO}) is decoupled so that one can obtain exactly the MI boundaries with $\mu^+=0.2U$, $\mu^-=0.0$. For other values of $t, \lambda$, the Hamiltonian is not exactly solvable even in one dimension, and hence we determine them numerically by using the DMRG techniques.
One can see that the MI region shrinks with increasing $\lambda$, although the phase boundary for each $\lambda$ looks similar.
This implies that the hopping integral $t$ is renormalized effectively larger for larger $\lambda$. Unfortunately,
this effect cannot sufficiently disclose internal structures in both MI and SF phases.

\begin{figure}[h]
\includegraphics[width=9cm, clip]{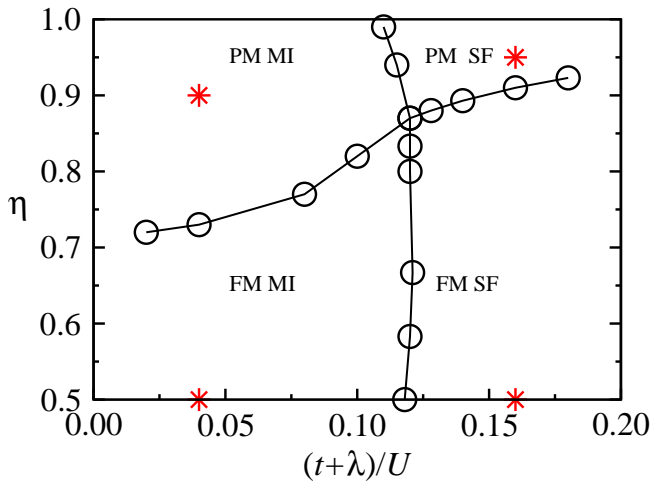}
\caption{(Color online) A $(t+\lambda, \eta)-$phase diagram with $U^{\prime} = 0.2U$.
The phase diagram shows only $\eta\in [0.5, 1]$ owing to its symmetry with respect
to the axis of $\eta=0.5$ (see text). A star stands for a representative point in each phase.}
\label{fig2}
\end{figure}

We note that the symmetry of the Hamiltonian (\ref{HSO}) is readily helpful in exploring the nature of the magnetic states in both MI and SF phases. TBHM possesses a U(1) $\times$ U(1) symmetry to conserve the particle number for each component. When ${\hat \mathcal{T}}_{SO}$ is added to TBHM, this symmetry is reduced to $U(1)\times Z_2$  described by the transformation
\begin{eqnarray}
\hat c_{i\tau^{\prime}} \rightarrow \sum_\tau [e^{i\theta} e^{-i\pi\sigma_y/2} ]_{\tau^{\prime} \tau} \hat c_{i\tau}.
\label{U1Z2}
\end{eqnarray}
In this case, only the total particle number is conserved. Moreover, Eq. (\ref{HSO}) is unchanged by interchanging $t$ and $\lambda$ with the following transformation
\begin{eqnarray}
&&\hat c_{i \tau}~~~\rightarrow ~~{\rm sgn}_\tau ~\hat c_{i \tau}, \ \  \ \ ~\hat c_{i+1 \tau} \rightarrow ~~\hat c_{i+1 \bar{\tau}}, \nonumber  \\
&&\hat c_{i+2 \tau} \rightarrow -{\rm sgn}_\tau ~\hat c_{i+2 \tau}, \ \ \hat c_{i+3 \tau} \rightarrow -\hat c_{i+3 \bar{\tau}}
\label{MAP}
\end{eqnarray}
for every 4-sites with ${\rm sgn}_\uparrow=1$ and ${\rm sgn}_\downarrow=-1$ and $\bar \tau$ represents the
opposite spin of $\tau$. It turns out that we can establish a $(t+\lambda, \eta)$-phase diagram  with  $\eta=\lambda/(t+\lambda)\in [0,1]$.
In Fig. \ref{fig2}, the phase diagram is presented only for the part of $0.5 \leq\eta\leq 1$, since the part of $0\leq\eta< 0.5$ is given
by interchanging $t$ and $\lambda$. One can see that new y-axis Ising ferromagnetic (FM) MI and SF phases appear in addition to paramagnetic (PM) MI and SF phases, respectively. These two FM phases essentially reveal the spontaneous breaking of the $Z_2$ symmetry and occur in the region where the SO coupling term $\hat {\mathcal T}_{SO}$ is comparable to the hopping kinetic energy, while two PM phases correspond to either $t\ll \lambda$ or $t \gg \lambda$.

\subsection{Criticality of the phase transitions}

In this subsection, we will discuss how to determine the phase boundary between PM phases and FM phases.
The PM phases are actually critical.
It turns out that it is rather difficult to determine accurately the critical value $\eta_c$ for the transition
numerically from the algebraically decayed correlation function (order parameter). Instead,
we determine $\eta_c$ by the entanglement entropy \cite{AMICO1} $S_{\rho}$,
which is defined as $S_{\rho} = -{\rm tr}\rho\ln\rho$ via the reduced density-matrix $\rho$ of a half-chain.
\begin{figure}[h]
\includegraphics[width=9cm, clip]{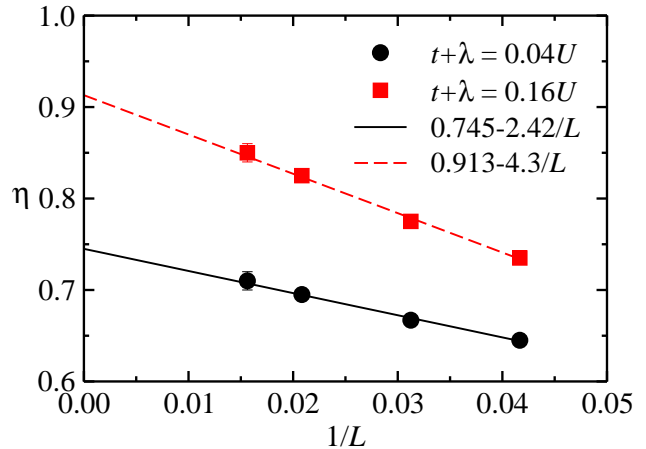}
\caption{(Color online) Finite-size scaling of the maximum of $S_{\rho}$ to determine the critical values. Symbols represent $\eta$ with maximum $S_{\rho}$ for the chain length $L=24,32,48,64$.}
\label{fig3}
\end{figure}
For this purpose, we first determine $\eta$ with maximum $S_{\rho}$ for each length, and then deduce the critical value of $\eta$ in the thermodynamic limit by making an extrapolation with respect to $1/L$.
Fig. \ref{fig3} shows such an extrapolation for different $\eta$ with chain length $L=24, 32, 48, 64$ at given $t+\lambda$  to determine $\eta_c$, which follows well a linear scaling behavior. We obtain $\eta_c = 0.745(20)$ at $t+\lambda=0.04U$ (in the MI phases) and $\eta_c = 0.913(20)$ at $t+\lambda=0.16U$ (in the SF phases). In the meanwhile, we also examine the energies of the ground and two excited states as a function of $\eta$ for each $t+\lambda$ and did not find level-crossing. Therefore the transition from the PM (MI or SF) phases to the FM (MI or SF) phases is continuous.
\begin{figure}
\includegraphics[width=9cm, clip]{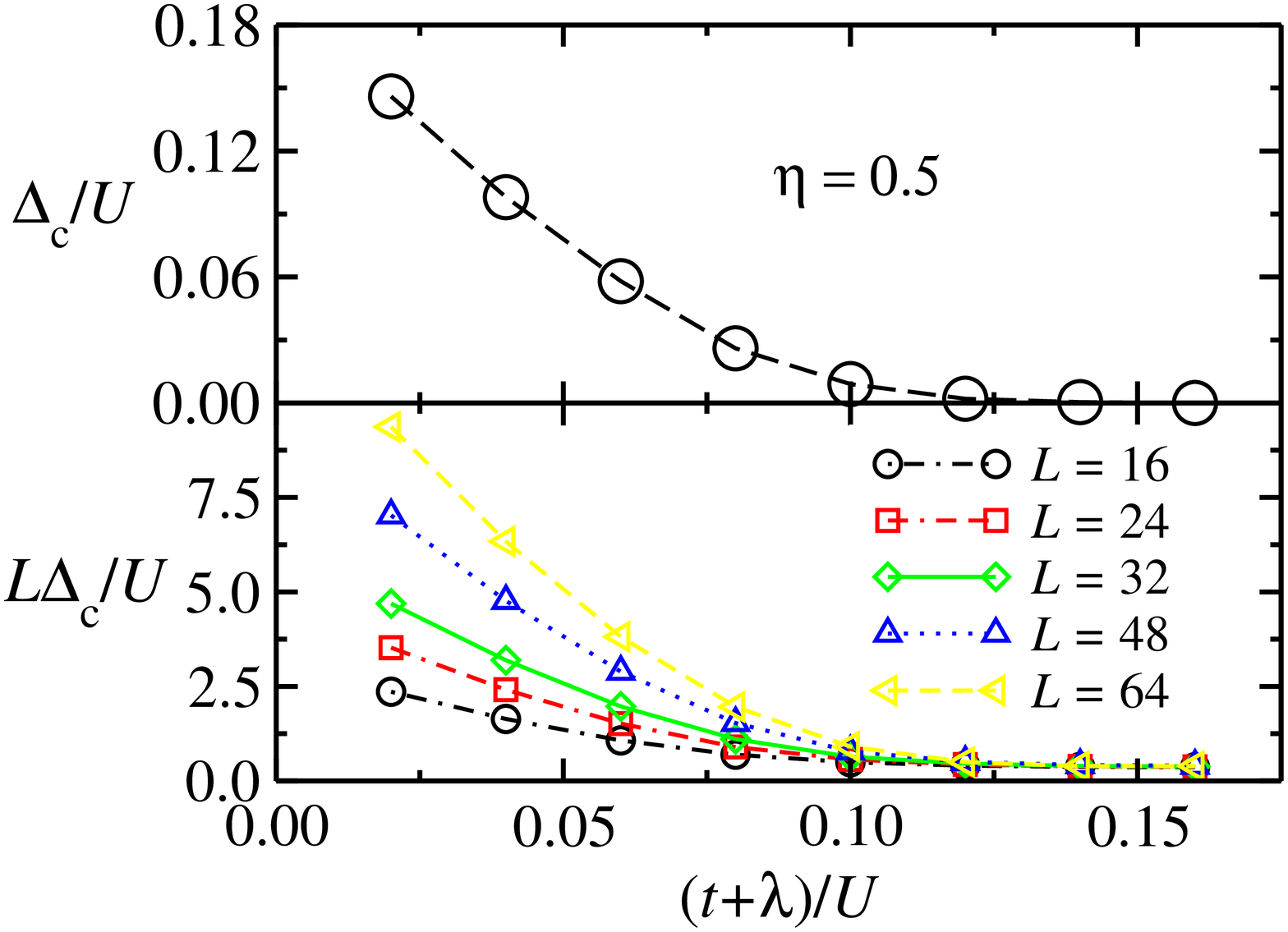}
\caption{(Color online) Determining the critical value of the BKT transition between the FM MI phase and the FM SF phase at $\eta=0.5$. Top panel shows the charge gap $\Delta_c$ as a function of $t+\lambda$. Bottom panel gives a finite-size scaling analysis of $\Delta_c$.
The critical value is estimated as $t+\lambda \simeq 0.12U$.}
\label{fig4}
\end{figure}
\begin{figure}[h]
\includegraphics[width=9cm, clip]{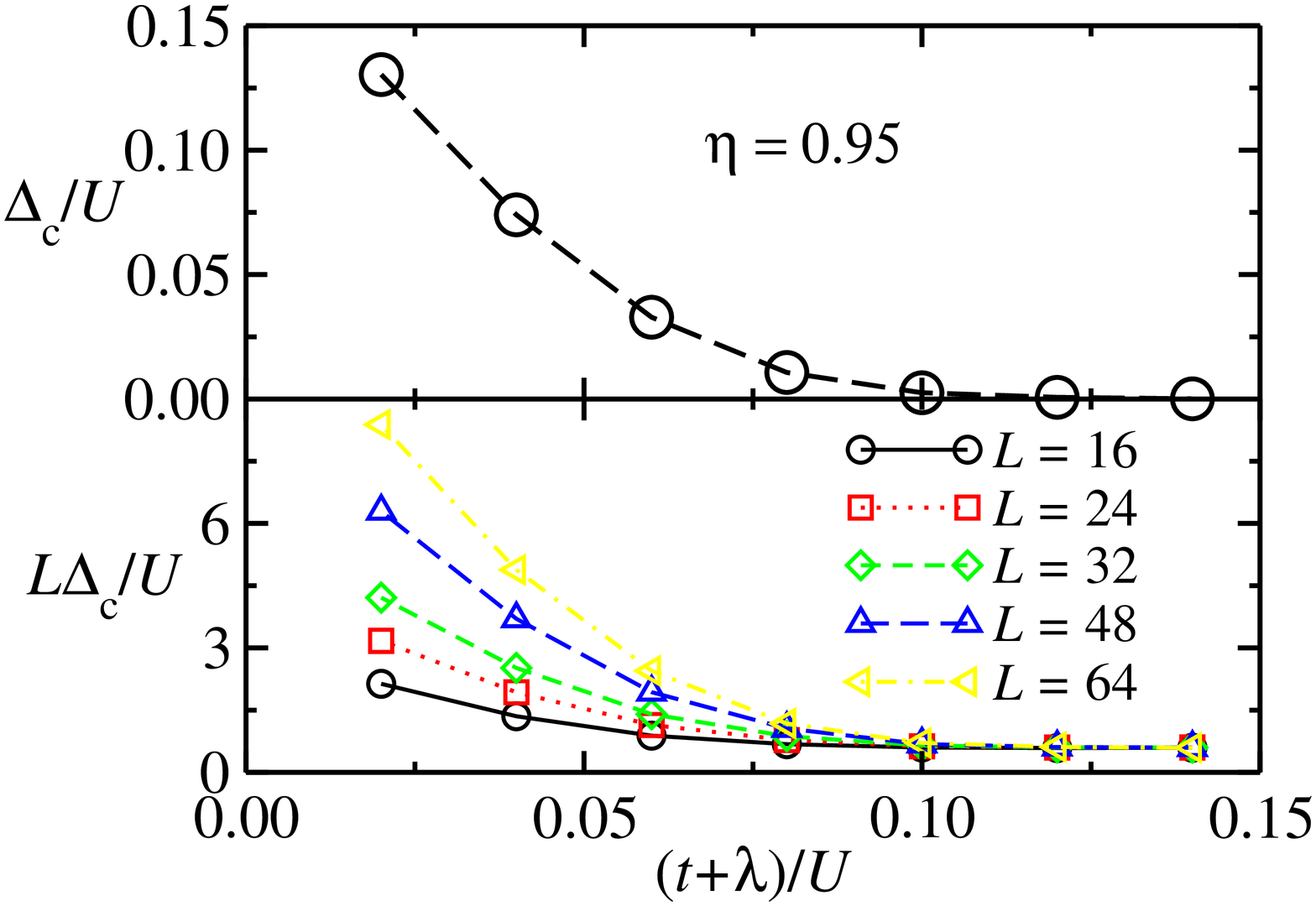}
\caption{(Color online) The same as Fig. \ref{fig4} but for $\eta=0.95$,  a transition from the PM MI phase to the PM SF phase. The critical value is estimated as  $t+\lambda\simeq 0.115U$.}
\label{fig5}
\end{figure}

Secondly we discuss the nature of the transition between the MI phases and the SF phases.
Since the boson density is fixed in our computation, a global density fluctuation is prohibited and only phase fluctuation is allowed. We expect that the Mott-superfluid transitions in  both PM phases and FM phases are of Berezinskii-Kosterlitz-Thouless(BKT)\cite{BKT} type. This can be verified by a finite-size scaling analysis\cite{PAI1, PAI2} of the charge gap $\Delta_c$.
Close to the critical point, the gap $\Delta_c$ scales as \cite{ROOMANY1}
\begin{eqnarray*}
L\Delta_c \sim f(L/\xi),
\end{eqnarray*}
where $\xi$ is the correlation length, which diverges in the superfluid phase.
For sufficiently large and different $L$, $L\Delta_c$ collapse into one curve in superfluid phase.
Fig. \ref{fig4} illustrates how to determine the transition point from a FM MI phase to FM SF phase at $\eta=0.5$. In the top panel, we show $\Delta_c$ at $L=\infty$ as a function of $t+\lambda$.
In the bottom panel, we show $L\Delta_c$ as a function of $t+\lambda$. With different chain lengthes $L$, $L\Delta_c$ collapse for sufficiently large values of $t+\lambda$, and divorces otherwise, showing the BKT type transition at $t+\lambda \simeq 0.12U$, in agreement with that obtained from the upper panel. By the same way, we found also the BKT type transition for the PM case. Fig. \ref{fig5} gives a critical value of $t+\lambda\simeq 0.115U$ at $\eta=0.95$.

\subsection{Excitation gaps and correlation functions}

In this subsection, we will interpret the nature of four different phases in Fig. \ref{fig2} in details. For our purpose, we take one point
in each phase and calculate several important quantities such as two kinds of excitation gaps, three different correlation functions. Four points (see stars in Fig. \ref{fig2}) correspond to $(t, \lambda)$ being $(0.004U, 0.036U)$ for PM MI, $(0.02U, 0.02U)$ for FM MI, $(0.008U, 0.152U)$ for PM SF and $(0.08U, 0.08U)$ for FM SF.

In addition to the single-particle excitation gap $\Delta_c$ for determining the boundaries of the MI phases of Fig.\ref{fig1}, we calculate longitudinal gaps $\Delta_l^k = E_k(N,L) - E_0(N,L)$ where $E_k(N,L)$ is the energy of the $k$th excited state with $N$ bosons for given $L$. We take $k = 1, 2$ to identify whether the ground states are two-fold degenerate in the MI-phases. The one-body density-matrix to distinguish SF states from  MI ones is defined as
\begin{equation}
n^\tau_{ij} = \langle\psi_0| \hat c^{\dagger}_{i \tau} \hat c_{j \tau}|\psi_0\rangle,
\end{equation}
and the spin-spin correlation function to justify possible magnetic orders is given by
\begin{equation}
\mathcal{S}^{\nu}_{ij} = \langle\psi_0|\hat S^{\nu}_{i} \hat S^{\nu}_{j}|\psi_0\rangle,
\end{equation}
where the spin operator is defined by $\hat S^{\nu} = \sum_{\tau, \tau'} \hat c^{\dagger}_\tau \sigma^{\nu}_{\tau \tau^\prime} \hat c_{\tau^\prime}/2$ with $\nu = x,y,z$,
and a particle-hole pairing correlation function to show peculiar feature of PM phases is defined as
\begin{equation}
{\mathcal C}_{ij} = \langle\psi_0|\hat c^\dagger_{i\uparrow}\hat c_{i\downarrow} \hat c_{j\uparrow} \hat c^\dagger_{j\downarrow}|\psi_0\rangle,
\end{equation}
for the ground state $|\psi_0\rangle$, and  $i$ being one of two central sites and $j$ in the system block when the DMRG sweeping process is completed \cite{WHITE1,HU2}.
\begin{figure}[h]
\includegraphics[width=9cm, clip]{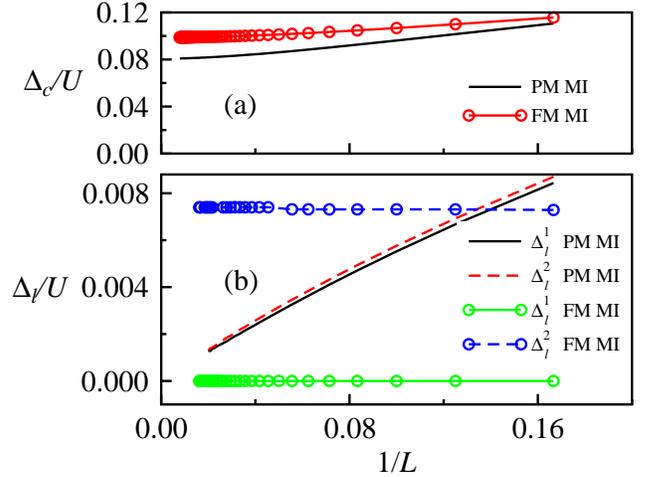}
  \caption{(Color online) Excitation gaps $\Delta_c$, $\Delta_l^1$ and $\Delta^2_l$ are shown as a function of $1/L$ for the
  representative points in two Mott insulator phases. Lines denote the results of FM MI with symbols, and those of PM MI without symbols otherwise.}
\label{fig6}
\end{figure}

\begin{figure}[h]
\includegraphics[width=9cm, clip]{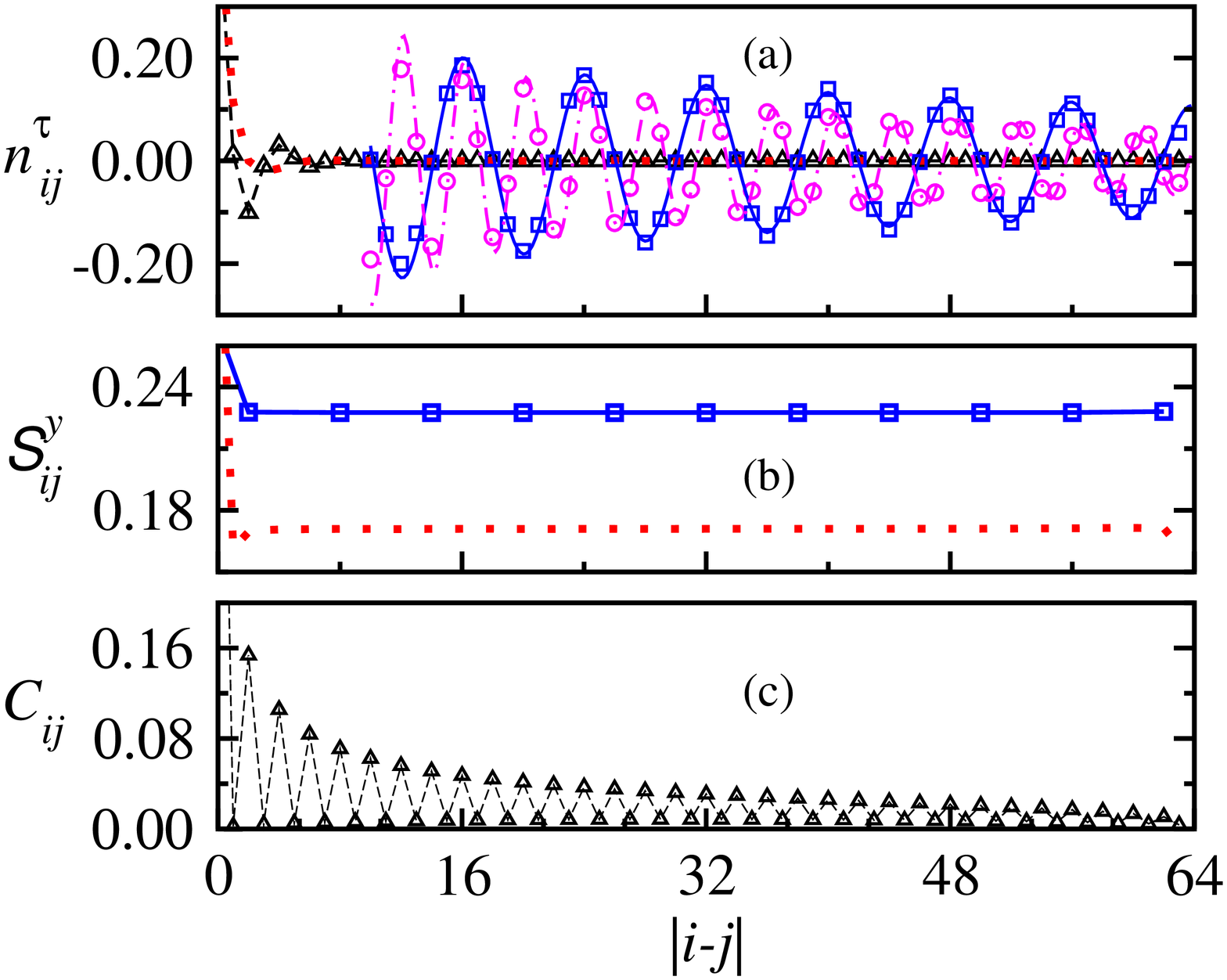}
\caption{(Color online) Characteristic correlation functions are shown as a function of $|i-j|$ in corresponding phases.
Black triangles for PM MI phase, red dots for FM MI phase, magenta circles for PM SF phase and blue squares for FM SF phase.
(a) $n^\tau_{ij}$ are shown for four phases. In two SF phases, $n^\tau_{ij}$ are shown for $|i-j|\geq 10$. Blue solid and magenta
dot-dashed lines fit numerical results (see text).
(b) $\mathcal S^y_{ij}$ are shown in both FM MI and FM SF phases. They are finite in the large $|i-j|$ limit.
(c) ${\mathcal C}_{ij}$ is shown for PM MI phase.  Algebraic dependence of ${\mathcal C}_{ij}$ on $|i-j|$
is found for the two PM phases, but almost shows no qualitative difference between such two phases
so that we displays ${\mathcal C}_{ij}$ only for PM MI phase.}
\label{fig7}
\end{figure}

Both PM MI and FM MI phases involve a finite $\Delta_c$ shown in Fig. \ref{fig6} (a) and the exponentially behavior of ${n}^{\tau}_{ij}$ shown in Fig. \ref{fig7}(a), as for the usual MI phase of TBHM. However, a quantum phase transition exists between PM MI and FM MI phases, which is driven by tuning the SO coupling. In PM MI phase, the SO coupling term $\hat{\mathcal T}_{SO}$ is dominant over the hopping kinetic energy. As aforementioned this region is equivalent to the one where the hopping kinetic energy is dominant over $\hat{\mathcal T}_{SO}$ via interchanging $t$ and $\lambda$. In this case, the low energy properties are qualitatively the same as those of TBHM such that the SO coupling effectively renormalizes the hopping integral. In particular, $\Delta^1_l=\Delta^2_l=0$ in the thermodynamic limit shown in Fig. \ref{fig6} (b) indicates the existence of a gapless mode in PM MI phase. This is verified by the algebraical behavior of $\mathcal C_{ij}$ with $|i-j|$ as shown in Fig. \ref{fig7} (c). The power-law behavior can be interpreted as a kind of counterflow superfluid of composite particles consisting of one particle from one component and one hole from the other component such that the net particle transfer is zero \cite{KUKLOV1}.
However, in FM MI phase, where $\hat{\mathcal T}_{SO}$ becomes comparable to the hopping kinetic energy, we found that the $Z_2$ symmetry is spontaneously broken, leading to a y-axis Ising FM long-range order confirmed by a finite $\mathcal S^y_{ij}$ shown in Fig. \ref{fig7}(b). Moreover, the ground state is two-fold degenerate and a finite $\Delta^2_l$ shown in Fig. \ref{fig6} (b) reveals gapful excitation despite of that the total $S^z$ is no longer a conserved quantity at $\lambda\neq 0$.

In both PM SF and FM SF phases of Fig. \ref{fig2}, we found that $\Delta_c=0$, $\Delta^1_l=\Delta^2_l=0$ and ${n}^{\tau}_{ij}\sim \cos((i-j)\pi/\varsigma-\delta)/|i-j|^\gamma$ for large $|i-j|$ as shown in Fig. \ref{fig7}(a).
Such pow-law behaviors mark the SF phases in one dimension,
and $\varsigma$, $\delta$ and $\gamma$ depend upon the values of $t/U$, $U^\prime/U$ and $\lambda/t$. For instances, with given $U^\prime=0.2U$, we obtain that $\varsigma=2.02$, $\delta=0.02\pi$ and $\gamma=0.8$ for $t=0.008U, \lambda=0.152U$, and $\varsigma=3.98$, $\delta=0.05\pi$ and $\gamma=0.45$ for $t=\lambda=0.08U$. Obviously, $n_{ij}^\tau$ behaves similarly in both FM SF and PM SF phases. The difference between these two phases is reflected by $\mathcal{S}^y_{ij}$. Only in FM SF phase, $\mathcal{S}^y_{ij}$ is finite for large $|i-j|$ as seen in Fig. \ref{fig7} (b) to display a long-range FM order along y-axis. This FM order also results from the spontaneous breaking of the $Z_2$ symmetry.
\begin{figure}[h]
\includegraphics[width=9cm, clip]{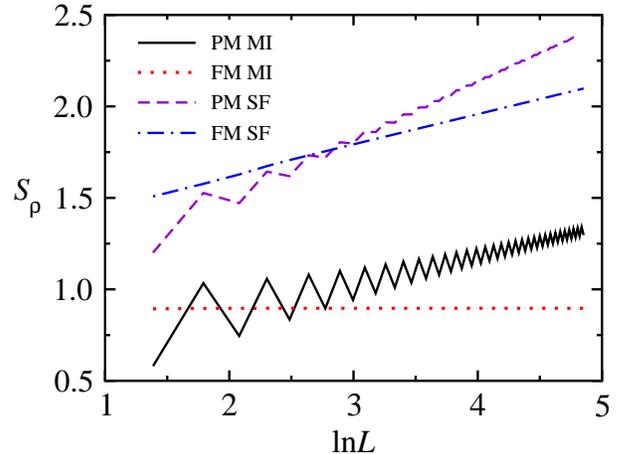}
\caption{(Color online) Entanglement entropy $S_\rho$ are shown as a function of $\ln{L}$ for four representative points
in different phases, from which the central charges are deduced with the logarithmic law Eq. (\ref{SRHO}) and given in Table I.}
\label{fig8}
\end{figure}

\subsection{Central charge}

We remark that low energy properties in both FM MI and FM SF phases are similar to those of a one-component system,
owing to the spontaneously breaking of the $Z_2$ symmetry. In this subsection, we will clarify this issue by the
entanglement entropy. According to conformal field theory, $S_{\rho}$  follows up a universal scaling law \cite{CALABRESE1}
\begin{equation}
S_{\rho} = \frac{c}{6}\ln{L} + C,
\label{SRHO}
\end{equation}
where $c$ is the central charge, a quantity that can indicate the number of gapless modes \cite{EISERT1} and $C$ is a nonuniversal constant. Fig. \ref{fig8} shows $S_{\rho}$ as a function of $\ln L$ for four representative points of Fig. \ref{fig2}. $S_{\rho}$ is ideally a straight line for both FM MI and FM SF phases so that one can easily extract the central charges based on Eq.(\ref{SRHO}). However, for either PM MI or PM SF phases, $S_{\rho}$ oscillates with respect to increasing $L$ and the oscillations reduce for large $|i-j|$. In this case, we use the central values of oscillating $S_\rho$ to fit Eq. (\ref{SRHO}). Up to the second digit accuracy, we obtain the central charges presented in Table I. One can see that two neighbor phases differ from each other
with different $c$. In the PM-MI and PM-SF phases, $c$ is found to be $1$ and $2$, respectively.
However, in the FM-MI phase $c=0$ and in the FM-SF phase $c=1$, which are in agreement with that of one-component Bose-Hubbard model. It is interesting to notice that the values of $c$ are consistent with what $\Delta_c$, $\Delta^2_l$ and $\mathcal S^y_{ij}$ show as seen in Table I, elegantly elaborating on the nature of the four phases.

\begin{table}[h]
\caption{The central charge $c$ extracted from the logarithmic law of $S_\rho$ in Fig. \ref{fig8}, the single particle excitation gap $\Delta_c$, the longitudinal gap $\Delta^2_l$, and the spin correlation function ${\mathcal S}^y_{ij}$ at large $|i-j|$ are listed together to further elucidate different natures of the four phases.}
\begin{tabular}{c|c|c|c|c}
\hline\hline
              &~PM-MI~& ~FM-MI~  & ~PM-SF~ & ~FM-SF~ \\ \hline
$c$& 1.0   & 0.0      & 2.0        &  1.0 \\ \hline
$~~\Delta_c~~$    & $>0$& $>0$   & 0        &  0 \\ \hline
$~\Delta^2_l~$    & 0   & $>0$   & 0        & $0$ \\ \hline
$~\mathcal S^y_{ij}$ &0 & $>0$ & 0         & $>0$\\  \hline\hline
\end{tabular}
\label{table1}
\end{table}

\subsection{Discussions}

Here we discuss relevant issues to experimental observation on the SOC effects for the present systems. 
For a general case of $U^\prime<U$, the ferromagnetic order emerges in both the MI and the SF phases 
when the kinetic energy and the SOC are comparable. 
To understand intuitively this phenomenon, we discuss the Hamiltonian (\ref{HSO}) in the deep Mott insulator region, 
where $t, \lambda \ll {U, U^\prime}$. In this region, Hamiltonian (\ref{HSO}) can be easily mapped to a spin-1/2 Heisenberg model. In particular, at $\lambda=0$, i.e. $\eta=0$ it is an XXZ Heisenberg model\cite{ALTMAN1} with its ground state in a critical phase. 
However, when $\lambda=t$, corresponding to $\eta=0.5$, the Hamiltonian reads
\begin{eqnarray}
\mathcal{H}^{eff} = \frac{-8t^2}{U^\prime}\sum_{i}\hat{S}^y_i\hat{S}^y_{i+1}-\frac{8t^2}{U}\sum_i\left(\hat{S}^z_i\hat{S}^x_{i+1}-\hat{S}^x_i\hat{S}^z_{i+1}\right),
\label{HEFF}
\end{eqnarray}
which can be further simplified into an exactly solvable XYX Heisenberg model\cite{Oshikawa}.
In the present case of $U^\prime<U$,  its ground state has a ferromagnetic long range order with 
a polarization along the $y$-axis. The FM order is stabilized by a finite gap so that it can exist 
in a finite range of $\eta$. Therefore, one can expect that a phase transition occurs when $\eta$ varies from 0 to 0.5. 
This is consistent with our numerical results.

In ultracold atomic experiments, a harmonic trap potential $V_i=\frac{1}{2}\omega^2\left(i-\frac{L}{2}\right)^2$ is necessary to load 
bosonic atoms into optical lattices. The effects of a weak trap potential can be understood approximately by local 
density approximation\cite{BERGKVIST1,BATROUNI1}. 
Under such approximation, the local density is determined by the local chemical potential $\mu_i=\mu-V_i$. By adjusting 
the interaction, proper Mott plateaus would emerge. Notice that the harmonic trap 
does not break the symmetries given by Eq. (\ref{U1Z2}) and (\ref{MAP}) and the FM order we predicted is protected by 
an excitation gap $\Delta_l^2$, therefore we can expect that the FM order is robust in the Mott region even in the presence of a 
trap potential. 
 
It is a highly nontrivial effect of the SOC that the FM phases result from the spontaneous breaking of 
the discrete $Z_2$ symmetry. This phenomenon actually cannot be ruled out by the Mermin-Wigner theorem 
for finite temperatures even in one dimension. In particular,
the finite excitation gap stabilizes such an FM order against not only quantum fluctuations but also thermal fluctuations. 
For instance, at $t=\lambda=0.02U$ (see Fig. \ref{fig6}), the 
FM order would be robust when a temperature is below the excitation gap $\Delta_l^2\approx t/3$ and is expected to be detectable via Bragg scattering of light\cite{CORCOVILOS1}. 

\section{Conclusions}

In conclusion, the phase diagram for one dimensional two-component bosonic Hubbard model with the synthetic
SO coupling is presented to show two ferromagnetic long-range order phases
in addition to the paramagnetic MI and SF phases. The nature of the four phases are well described
in terms of excitation gaps, the correlation functions and the entanglement entropy.
Since the new phases are inherently related to the spontaneous breaking of the $Z_2$ symmetry, but occur
only when the SO coupling becomes comparable to the hopping kinetic energy and the intercomponent interaction
smaller than the intracomponent one as well, they should also occur in two dimensions and for continuous models.
Moreover, experimental setups are expected to detect such intrinsic effects of the SO coupling in the near future since SO coupling
has been realized in experiments\cite{LIN1}, and two-component bosonic systems have
already been loaded into a one-dimensional optical lattice\cite{FUKUHARA1}.

\section{Acknowledgements}

We thank Peng Zhang for fruitful discussions. The computational resources are provided by the high-performance computer-Kohn at physics department, Renmin University of China.
X.Q. Wang is supported by MOST 2012CB921704 and NSFC 11174363, P. Zhang is supported by NSFC 90921003.

\vfill

\begin{references}
\bibitem{ANDERSON1} M.H. Anderson, J.R. Ensher, M.R. Matthews, C.E. Wieman, and E.A. Cornell, Science {\bf{269}}, 198 (1995).
\bibitem{DAVIS1} K.B. Davis, M.O. Mewes, M.R. Andrews, N.J. van Druten, D.S. Durfee, D.M. Kurn, and W. Ketterle, Phys. Rev. Lett. {\bf{75}}, 3969 (1995).
\bibitem{FISHER1} M.P.A. Fisher, P.B. Weichman, G. Grinstein, and D.S. Fisher, Phys. Rev. B {\bf{40}}, 546 (1989).
\bibitem{DUAN1} L.M. Duan, E. Demler, and M.D. Lukin, Phys. Rev. Lett. {\bf{91}}, 090402 (2003).
\bibitem{FUKUHARA1} T. Fukuhara, P. Schau$\ss$, M. Endres, S. Hild, M. Cheneau, I. Bloch, and C. Gross, Nature, {\bf{502}}, 76 (2013).
\bibitem{GREINER1} M. Greiner, O. Mandel, T. Esslinger, T.W. Hansch, and I. Bloch, Nature {\bf{415}}, 39 (2002).
\bibitem{LIN1} Y.J. Lin, K. Jimenez-Garcia, and I.B. Spielman, Nature {\bf{471}}, 83 (2011).
\bibitem{DZ}I. Dzyaloshinsky, J. Phys. and Chem. Sol. {\bf 4}, 241(1958).
\bibitem{MORIYA}T. Moriya, Phys. Rev. {\bf 120}, 91(1960).
\bibitem{Broholm} D.C. Dender, P.R. Hammar, D.H. Reich, C. Broholm, and G. Aeppli, Phys. Rev. Lett. 79, 1750 (1997).
\bibitem{Oshikawa} M. Oshikawa and I. Affleck, Phys. Rev. Lett. 79, 2883 (1997).
\bibitem{Zhao} J.Z. Zhao, X.Q. Wang, T. Xiang, Z. B. Su, and L. Yu£¬Phys. Rev. Lett. 90, 207204 (2003).
\bibitem{KANE1} C.L. Kane and E.J. Mele, Phys. Rev. Lett. {\bf{95}}, 226801 (2005).
\bibitem{QI1} X.L. Qi, T.L. Hughes, and S.C. Zhang, Phys. Rev. B {\bf{78}}, 195424 (2008).
\bibitem{WANG1} P.J. Wang, Z.Q. Yu, Z.K. Fu, J. Miao, L.H. Huang, S.J. Chai, H. Zhai, and J. Zhang, Phys. Rev. Lett. {\bf{109}}, 095301 (2012).
\bibitem{CHEUK1} L.W. Cheuk, A.T. Sommer, Z. Hadzibabic, T. Yefsah, W.S. Bakr, and M.W. Zwierlein, Phys. Rev. Lett. {\bf{109}}, 095302 (2012).
\bibitem{WU1} Congjun Wu , Ian Mondragon Shem, and Xiang-Fa Zhou, Chin. Phys. Lett. {\bf{28}}, 097102 (2011).
\bibitem{ZHOU1} Xiang-Fa Zhou, Jing Zhou, Congjun Wu, Phys. Rev. A {\bf{84}}, 063624 (2011).
\bibitem{HU4} Hui Hu, B. Ramachandhran, Han Pu, and Xia-Ji Liu, Phys. Rev. Lett. {\bf{108}}, 010402 (2012).
\bibitem{DENG1} Y. Deng, J. Cheng, H. Jing, C.-P Sun, and S. Yi, Phys. Rev. Lett. {\bf{108}}, 125301 (2012).
\bibitem{ZHANG1} Wei Zhang and Wei Yi, Nat. Communications {\bf{4}}, 2711 (2013).
\bibitem{QU1} Chunlei Qu, Zhen Zheng, Ming Gong, Yong Xu, Li Mao, Xubo Zou, Guangcan Guo, and Chuanwei Zhang, Nat. Communications {\bf{4}}, 2710 (2013).
\bibitem{ZHAI1} H. Zhai, Int. J. Mod. Phys. B {\bf{26}} 1230001 (2012).
\bibitem{CAI1} Z. Cai, X.F. Zhou, and C.J. Wu, Phys. Rev. A  {\bf{85}}, 061605 (2012).
\bibitem{MANDAL1} S. Mandal, K. Saha, and K. Sengupta, Phys. Rev. B {\bf{86}}, 155101 (2012).
\bibitem{RADIC1} J. Radic, A. DiCiolo, K. Sun, and V. Galitski, Phys. Rev. Lett. {\bf{109}}, 085303 (2012).
\bibitem{GONG1} M. Gong, Y.Y. Qian, V.W. Scarola, and C.W. Zhang, arXiv: 1205.6211.
\bibitem{COLE}W.S. Cole, S.Z. Zhang, A. Paramekanti, and N. Trivedi, Phys. Rev. Lett. {\bf{109}}, 085302 (2012).
\bibitem{ALTMAN1} E. Altman, W. Hofstetter, E. Demler, and M. D Lukin, New J. Phys. {\bf{5}} 113 (2003).
\bibitem{MISHRA1} T. Mishra, R.V. Pai, and B.P. Das, Phys. Rev. A {\bf{76}}, 013604 (2007).
\bibitem{HU3} A. Hu, L. Mathey, I. Danshita, E. Tiesinga, C.J. Williams, and C.W. Clark, Phys. Rev. A {\bf{80}}, 023619 (2009).
\bibitem{ISACSSON1} A. Isacsson, M.C. Cha, K. Sengupta, and S.M. Girvin, Phys. Rev. B {\bf{72}}, 184507 (2005).
\bibitem{WHITE1} S.R. White, Phys. Rev. Lett. {\bf{69}}, 2863 (1992).
\bibitem{PESCHEL}I. Peschel, X.Q. Wang, M. Kaulke, and K. Hallberg, Density Matrix Renormalization, LNP528 (1999), Springer.
\bibitem{SCHOLLWOCK1} U. Schollw$\ddot{o}$ck, Rev. Mod. Phys. {\bf{77}}, 259 (2005).
\bibitem{HU1} S.J. Hu, Y.C. Wen, Y. Yu, B. Normand, and X.Q. Wang, Phys. Rev. A 80, 063624 (2009).
\bibitem{HU2}S.J. Hu, B. Normand, X.Q. Wang, and Lu Yu, Phys. Rev. B 84, 220402 (2011).
\bibitem{KUHNER1} T. D. K\"{u}hner and H. Monien, Phys. Rev. B {\bf{58}}, R14741 (1998).
\bibitem{AMICO1} L. Amico, R. Fazio, A. Osterloh, and V. Vedral, Rev. Mod. Phys. {\bf{80}}, 517 (2008).
\bibitem{BKT} V. L. Berezinskii, Sov. Phys. JETP {\bf{32}}, 493 (1971); J. M. Kosterlitz and D. J. Thouless, J. Phys. C: Solid State Phys. {\bf{6}}, 1181 (1973).
\bibitem{PAI1} R. V. Pai, R. Pandit, H. R. Krishnamurthy, and S. Ramasesha, Phys. Rev. Lett. {\bf{76}}, 2937 (1996).
\bibitem{PAI2} R. V. Pai and R. P. Pandit, Phys. Rev. B {\bf{71}}, 104508 (2005).
\bibitem{ROOMANY1} H. H. Roomany and H. W. Wyld, Phys. Rev. D {\bf{21}}, 3341 (1980).
\bibitem{KUKLOV1} A.B. Kuklov and B.V. Svistunov, Phys. Rev. Lett. {\bf{90}}, 100401 (2003);  A. Kuklov, N. Prokof'ev, and B. Svistunov, Phys. Rev. Lett. {\bf{92}}, 030403 (2004).
\bibitem{CALABRESE1} P. Calabrese and J. Cardy, J. Phys. A {\bf{42}}, 504005 (2009).
\bibitem{EISERT1} J. Eisert, M. Cramer, and M. B. Plenio, Rev. Mod. Phys. {\bf{82}}, 277 (2010).
\bibitem{BERGKVIST1} Sara Bergkvist, Patrik Henelius, and Anders Rosengren, Phys. Rev. A {\bf{70}}, 053601 (2004).
\bibitem{BATROUNI1} G. G. Batrouni, H. R. Krishnamurthy, K. W. Mahmud, V. G. Rousseau, and R. T. Scalettar, Phys. Rev. A {\bf{78}}, 023627 (2008).
\bibitem{CORCOVILOS1} T. A. Corcovilos, S. K. Baur, J. M. Hitchcock, E. J. Mueller, and R. G. Hulet, Phys. Rev. A 81, 013415 (2010).
\end{references}
\end{document}